%% file: heavy.tex
\documentclass[11pt,a4paper]{article}
\pdfoutput=1
\usepackage{graphicx,caption}
\usepackage{ifpdf}
\usepackage{xspace}
\usepackage{jheppub}
\usepackage{dsfont}
\usepackage{amsfonts}
\usepackage{amsmath}
\usepackage{amssymb}
\usepackage{bbold}
\usepackage{xurl}
\usepackage{hyperref}
\usepackage{axodraw2}
\usepackage{pstricks}
\usepackage{relsize,exscale,scalefnt,anyfontsize,cases}
\usepackage[utf8]{inputenc}
\include{macro}

\newcommand{\figscale}{.65}  

\title{Impact of fully coherent energy loss on heavy meson production in pA collisions}

\author[a,b]{Fran\c{c}ois Arleo,}
\author[c]{Greg Jackson,}
\author[b]{St\'ephane Peign\'e}

\affiliation[a]{Laboratoire Leprince-Ringuet, \'Ecole polytechnique, Institut polytechnique de Paris, CNRS/IN2P3, Route de Saclay, 91128 Palaiseau, France}
\affiliation[b]{SUBATECH UMR 6457 (IMT Atlantique, Universit\'e de Nantes, IN2P3/CNRS), 4 rue Alfred Kastler, 44307 Nantes, France}
\affiliation[c]{Institute for Nuclear Theory, Box 351550, University of Washington, Seattle, WA 98195-1550, United States}

\emailAdd{francois.arleo@cern.ch}
\emailAdd{gsj6@uw.edu}
\emailAdd{peigne@subatech.in2p3.fr}

\abstract{
  Hadron production in proton-nucleus (pA) collisions was previously shown to be suppressed 
  by medium-induced fully coherent energy loss (FCEL). 
  We show that the quenching of $D$ and $B$ mesons in pPb collisions at the LHC due solely 
  to FCEL is, at least, on par with other nuclear effects such as gluon shadowing or saturation. 
  This is consistent with previous findings for both quarkonium and light hadron production 
  in pA collisions, emphasising that FCEL effects need to be included for a reliable understanding 
  of hadron production measurements in pA collisions.
}

\keywords{perturbative QCD; proton--nucleus collisions; parton energy loss.}

\begin{document} 

\maketitle
\setcounter{footnote}{0}
\renewcommand{\thefootnote}{\arabic{footnote}}

\newpage

\section{Introduction}

Several nuclear effects influence the production of hadrons 
in high-energy proton-nucleus (pA) collisions, 
as compared to proton-proton (pp) collisions. 
For instance,
nuclear parton distribution functions (nPDFs) are known to differ from PDFs in a proton 
at all values of Bjorken-$x$ (see Ref.~\cite{Armesto:2006ph} for a review). 
In particular, the effect of gluon shadowing, namely, 
the depletion at $x \lesssim 10^{-2}$ of the gluon nuclear PDF with respect to that in a proton, 
leads to a corresponding suppression of hadron production in pA with respect to pp collisions, 
either at RHIC (at forward rapidity) or at LHC~\cite{Helenius:2012wd,QuirogaArias:2010wh}. 
As in the proton case, nPDFs are obtained from global fits based on 
DGLAP evolution~\cite{deFlorian:2011fp,Kovarik:2015cma,Eskola:2016oht,AbdulKhalek:2019mzd}, 
assuming collinear factorization~\cite{Collins:1989gx} to also hold in nuclear collisions. 
But nPDFs suffer from rather large theoretical uncertainties, especially at small $x$, 
due to the relative scarcity of data included in those analyses. 
Thus, the actual quantitative role of gluon shadowing is still being discussed.

The formalism that defines gluon saturation (see~\cite{Gelis:2010nm} for a review) 
incorporates additional effects when compared to gluon shadowing, 
in particular through the use of nuclear $k_{_\perp}\!$-dependent gluon 
distributions~\cite{Albacete:2012xq}. 
Original calculations overpredicted the nuclear suppression of light 
hadron~\cite{Albacete:2010bs} and quarkonium~\cite{Fujii:2013gxa} production at the LHC, 
but later revisions (see~\cite{Tribedy:2011aa,Albacete:2012xq,Rezaeian:2012ye,Lappi:2013zma} 
for light hadron and~\cite{Ducloue:2015gfa} for quarkonium production) proved to be consistent, 
within theoretical and experimental uncertainties, with LHC pPb data. 
Other nuclear effects such as $\pt\!$-broadening~\cite{Kopeliovich:2005ym} or initial-state 
parton energy loss~\cite{Frankfurt:2007rn,Kang:2012kc}, have also been considered in 
studies of hadron production in pA collisions at RHIC and LHC. 

Another important nuclear effect, {\it fully coherent energy loss} (FCEL) in cold nuclear matter, 
is expected in hadron and jet production in pA collisions, 
for which the underlying partonic process consists in forward scattering 
(when viewed in the target nucleus rest frame) of an incoming high-energy parton to an outgoing 
colour charge~\cite{Arleo:2010rb,Peigne:2014uha} or colourful system of partons~\cite{Peigne:2014rka}. 
The average energy loss, in these situations, is proportional to the energy $E$ of the incoming parton, 
$\Delta E_{_{\textnormal{FCEL}}} \propto E$~\cite{Arleo:2010rb}, 
thus overwhelming parton energy loss in the Landau--Pomeranchuk--Migdal regime which has 
milder $E$-dependence \cite{Baier:1996sk,Baier:1996kr,Zakharov:1996fv,Zakharov:1997uu}. 

FCEL has been computed from first principles in various 
formalisms~\cite{Arleo:2010rb,Armesto:2012qa,Armesto:2013fca,Peigne:2014uha,Peigne:2014rka,Liou:2014rha,Munier:2016oih}, 
and has proven to be crucial in understanding $J/\psi$ (and $\Upsilon$) nuclear suppression, 
from fixed-target energies (where FCEL alone can describe the world data on $J/\psi$ suppression) 
to collider energies~\cite{Arleo:2012hn,Arleo:2012rs,Arleo:2013zua}.\footnote{\label{foot:nPDFvsFCEL}%
  Let us stress that nPDF/saturation effects alone would not allow for such a global description 
  of the $J/\psi$ suppression data. 
  Indeed, those effects typically scale in $x_2$ (they tend to be sizable at collider energies, 
  but absent or minor at fixed-target energies), 
  but such a scaling is strongly violated in the $J/\psi$ suppression data~\cite{Hoyer:1990us} 
  (see also Fig.~5 of~\cite{Arleo:2018zjw}). 
  In contrast, FCEL has an approximate scaling in $x_1\,$, 
  and the resulting extrapolation of the FCEL effect from fixed-target to collider energies 
  allowed to successfully predict $J/\psi$ suppression at RHIC and LHC~\cite{Arleo:2012rs}. 
  There is certainly some room left for nPDF/saturation effects at those energies, 
  but the possibility for these effects to be the only ones at work at LHC seems very unlikely.
} 
More recently, the effect of FCEL on light hadron production in pPb collisions at the LHC has 
been studied~\cite{Arleo:2020eia,Arleo:2020hat}, and found to be quantitatively as important 
as gluon shadowing~\cite{Helenius:2012wd} or 
saturation~\cite{Tribedy:2011aa,Albacete:2012xq,Rezaeian:2012ye,Lappi:2013zma}. 
Thus, the proposal to use light hadron production data in pA collisions to better constrain 
nPDFs~\cite{Helenius:2012wd,QuirogaArias:2010wh} and saturation effects~\cite{Albacete:2012xq} 
should be followed cautiously, and by not discounting other known physical effects like FCEL.

Our first goal in this article is to recall that FCEL influences the production of {\it any} 
hadron in pA collisions~\cite{Arleo:2010rb}. 
Following on from what has been done for quarkonium and light hadron production, 
we will present baseline predictions for the nuclear suppression of $D$ and $B$ mesons 
expected from the sole FCEL effect. 
Isolating the role played by FCEL is motivated by its modest theoretical uncertainty, 
a virtue of being fully determined within perturbative QCD (pQCD).
Because future nPDF global fit analyses could benefit from incorporating FCEL, 
it is natural to study FCEL separately and before combining it with other effects 
that are hampered by larger uncertainties.

A second aim is to demonstrate that the FCEL effect on open heavy-flavour production is 
quantitatively sizable (as is the case for quarkonium~\cite{Arleo:2012hn,Arleo:2012rs,Arleo:2013zua} 
and light hadron production~\cite{Arleo:2020eia,Arleo:2020hat}), 
and turns out to explain about {\it half} of the heavy-flavour nuclear suppression 
observed at the LHC at forward rapidities ($2 < y < 4$) and for $\pt \lsim 5$ GeV. 
We also emphasise that with increasing $\pt$, 
the magnitude of the FCEL effect decreases faster than that of nPDF effects, 
although the contribution of FCEL to nuclear suppression remains significant even for $\pt \sim 10$~GeV. 

Lastly, since FCEL arises from first principles and is comparable in magnitude with nPDF effects, 
we argue that it should be taken into account in nPDF global fit analyses using the pA data on 
$D$/$B$ meson production (as well as on quarkonium and light hadron production). 
Hadron production in pA collisions at collider energies is frequently addressed assuming that 
nPDF effects are the only nuclear effect at work -- an assumption made for the sake of simplicity 
and only justified {\em a posteriori}, based on the goodness of global fits to world data. 
In particular, recent studies~\cite{Kusina:2017gkz,Eskola:2019bgf,Kusina:2020dki} proposed to consider, 
within the latter paradigm, 
the data on $D$ and $B$ meson production in pA collisions as a reliable probe of nPDFs. 
Our results suggest that such analyses are not exempt from FCEL.

The outline of the paper is as follows. 
In Sec.~\ref{sec:model} we present the physical picture 
used to implement FCEL in heavy-flavour production. 
Baseline calculations of FCEL effects on $D$ and $B$ nuclear suppression are discussed and 
compared to experimental data in Sec.~\ref{sec:predictions}. 
We conclude with a critical discussion in Sec.~\ref{sec:discussion}.

\section{Heavy-flavour production: physical picture}
\label{sec:model}

In this section we present the model used to single out the FCEL effect in 
heavy-flavour production in \pA collisions. 
The physical picture is the same as that used for quarkonium production 
(in pA~\cite{Arleo:2012hn,Arleo:2012rs,Arleo:2013zua} and heavy-ion~\cite{Arleo:2014oha} collisions), 
and for light hadron production~\cite{Arleo:2020eia,Arleo:2020hat}. 
The essential aspect common to these studies is the scaling of the FCEL quenching weight 
(defined below) in the fractional energy loss $x \equiv \varepsilon/E$, 
independently of the partonic process where FCEL occurs. 

\subsection{Subprocess and kinematics}
\label{sec:setup}

Denoting the produced heavy meson transverse momentum by $\pt$ and its rapidity by $y$ 
(in the c.m.~frame of an elementary proton--nucleon collision of energy $\sqrt{s}$), 
we focus on the kinematical domain of moderate $\pt \sim \morder{m}$ (with $m$ the heavy quark mass), 
and mid to large rapidities $|y| \leq 5$. 

\begin{figure}[t]
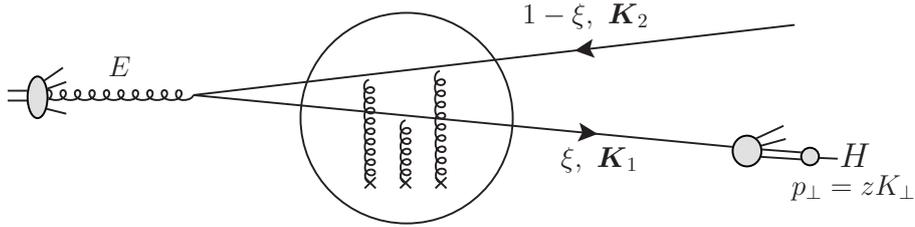

\begin{center}
\hskip -12mm \GenericGtoQQbar (115,-30,15) 
\end{center}
\caption{
  Contribution to the production of a heavy meson $H$ in pA collisions from the 
  LO process $gg \to Q \bar{Q}$ 
  (viewed as $g \to Q \bar{Q}$ forward scattering in the target nucleus rest frame), 
  followed by quark fragmentation $Q \to H$ (represented here), or $\bar{Q} \to H$. 
  (In addition to the incoming target gluon, 
  the nucleus provides rescattering gluons that contribute to the nuclear 
  $\pt\!$-broadening $\ell_{_{\perp \rm A}}$.) 
}
\label{fig:gtoQbarQ}
\end{figure}

In a leading-order (LO) pQCD picture, heavy-flavour production in pp and pA collisions 
at the LHC proceeds dominantly via the $gg \to Q \bar{Q}$ partonic reaction. 
The $q \bar{q} \to Q \bar{Q}$ process is indeed quite insignificant, 
due to the smallness of the antiquark PDF compared with that of the gluon. 
As for the $Q g \to Q g$ process, where the initial heavy quark $Q$ arises from (perturbative) 
gluon splitting $g \to Q \bar{Q}$, we choose to interpret it in the fixed flavour number 
scheme as a next-to-leading-order (NLO) `flavour-excitation' process~\cite{Mangano:1998oia}, 
rather than an LO process involving an input heavy quark PDF 
(as in the variable flavour number scheme~\cite{Buza:1996wv}). 
Although at moderate $\pt$ both interpretations should be equally valid for sufficiently 
inclusive cross sections~\cite{Mangano:1998oia}, 
interpreting $Q g \to Q g$ as part of the NLO process $gg \to Q \bar{Q} g$ allows to 
keep track of the produced heavy antiquark, 
and thus of the global colour state of the final parton system, 
which is more appropriate for our purpose (see Sec.~\ref{sec:rescaling}). 
With this choice, the only important process to consider at LO is thus $gg \to Q \bar{Q}$.\footnote{
  Let us remark that in the case $Q=c$ ($Q=b$), 
  the $Q g \to Q g$ process might contribute via some non-perturbative intrinsic charm 
  (bottom) component in the projectile proton. 
  However, independently of its magnitude (expected to be quite small already for intrinsic charm), 
  such a contribution should only play a role at {\it very large} rapidities, 
  close to the proton beam rapidity $y_{\mathrm{beam}} \simeq 8.5$ (for $\sqrt{s} = 5\,{\rm TeV}$).
}
When viewed in the target rest frame, 
this process looks like $g \to Q \bar{Q}$ forward scattering, as illustrated in Fig.~\ref{fig:gtoQbarQ}. 

The final heavy quarks $Q$ and $\bar{Q}$ have energy fractions $\xi$ and $1-\xi$ with 
respect to the incoming gluon energy $E$ (in the target rest frame), 
and transverse momenta $\Kvec_1$ and $\Kvec_2$, respectively. 
We will assume the transverse momentum imbalance $\Kvec_1 + \Kvec_2$ of the $Q \bar{Q}$ pair to be small, 
$|\Kvec_1 +\Kvec_2| \ll |\Kvec_1| \equiv K_{_\perp} \simeq |\Kvec_2|$. 
The $g \to Q \bar{Q}$ forward process is followed by quasi-collinear fragmentation of the heavy quark 
(or antiquark) into the tagged heavy meson $H$, 
which thus inherits the transverse momentum $\pt = z K_{_\perp}$, 
where $z$ is the fragmentation variable. 

We will use the above setup, where the heavy quarks are produced approximately 
back-to-back in the transverse plane, in both pp and pA collisions. 
We thus assume the gluon nuclear $\pt\!$-broadening $\ell_{_{\perp \rm A}}$ 
(or gluon saturation scale $Q_s$), defined below in \eq{gluon-broad}, 
to be smaller than the `hard scale' $K_{_\perp}$ of the subprocess, 
$\ell_{_{\perp \rm A}} \ll  K_{_\perp}$. 
This setup allows one to single out the quantitative effect of FCEL, 
independently of other existing nuclear mechanisms such as modified PDFs 
or the Cronin effect.\footnote{%
  The role of $\pt\!$-broadening in our study is simply to specify the input quantity 
  $\ell_{_{\perp \rm A}}$ entering the FCEL quenching weight $\Phat_{_\R}$, 
  see Sec.~\ref{sec:rescaling}.
}

To prepare for what comes next, 
let us quote the expressions of the heavy quark transverse mass $m_{_\perp}$, 
the $Q \bar{Q}$ pair rapidity difference $\Delta y \equiv y_{_Q}- y_{_{\bar{Q}}}$ and 
invariant mass $M_{\xi}$ in the above setup,
\be
\label{dijet-mass}
m_{_\perp}^2 \equiv K_{_\perp}^2 + m^2 \ \ ;  
\ \  \Delta y  = \log{\left(\frac{\xi}{1-\xi} \right)} \ \ ;
\ \  M_{\xi}^2 = \frac{m_{_\perp}^2}{\xi(1-\xi)} \, .
\ee

\subsection{Implementing FCEL}
\label{sec:rescaling}

Similarly to the study of light hadron production~\cite{Arleo:2020eia,Arleo:2020hat}, 
we will implement FCEL in the simplifying approximation where the $Q \bar{Q}$ pair behaves 
as a pointlike object with regard to the medium-induced radiation. 
This `pointlike dijet approximation' (PDA) holds when the induced radiation, 
of energy $\omega$ and transverse momentum $k_{_\perp}$, does not probe the size of the parton pair 
(at the time $\tf \sim \omega/k_{_\perp}^2$ of the induced radiation), 
nor the individual colour charges of the pair constituents. 
Similarly to the massless case~\cite{Arleo:2020hat}, 
one finds that those conditions are satisfied within the logarithmic accuracy 
\be
\label{log-accuracy}
\ln{\left(\frac{E^2 \ell_{_{\perp \rm A}}^2}{\omega^2 m_{_\perp}^2} \right)} \gg 1 \, ,
\ee
which thus defines the PDA.

In the PDA the induced spectrum $\omega \, {\dd{I}_{_\R}}/{\dd \omega}$ for a parton pair (or `dijet') 
in colour state $\R$ is directly obtained from the spectrum associated to a pointlike colour charge 
(derived in \cite{Arleo:2010rb,Arleo:2012rs,Peigne:2014uha,Munier:2016oih}) 
by replacing the particle's mass by the dijet mass 
given in Eq.~\eq{dijet-mass}, namely,
\be
\omega \frac{\dd I_{_\R}}{\dd \omega}  
=  \, (C_a + C_\R - C_b) \, \frac{\alpha_s}{\pi} 
\, \left\{ \ln{\left(1+\frac{E^2  \ell_{_{\perp\rm A}}^2}{\omega^2 M_{\xi}^2}\right)} - 
\ln{\left(1+\frac{E^2  \ell_{_{\perp \rm p}}^2}{\omega^2 M_{\xi}^2}\right)} \right\} \, , 
\label{dIR} 
\ee
where $C_a$ and $C_b$ are the (Casimir) colour charges of the incoming partons 
from the proton and target nucleus, respectively. 
For the present purpose, we focus on $gg \to Q \bar{Q}$ (see Sec.~\ref{sec:setup}), 
so that $C_a = C_b =N_c$, and $C_\R=N_c$ or $C_\R=0$ (in which case the FCEL spectrum \eq{dIR} vanishes) 
for the two possible colour states, respectively octet and singlet, of the $Q \bar{Q}$ pair. 
The coupling constant $\alpha_s$ in \eq{dIR} should be evaluated at the semi-hard scale 
$\ell_{_{\perp\rm A}} \sim \morder{1\, {\rm GeV}}$, where it will be assumed to be frozen, 
$\alpha_s = 0.5\,$. 

The quenching weight associated with FCEL is defined as a function of the medium-induced 
energy loss $\varepsilon$ by~\cite{Arleo:2012rs}
\be
\label{qw-R}
{\cal P}_{_\R}(\varepsilon, E)  = 
\frac{\dd I_{_\R}}{\dd\varepsilon} \, 
\exp \left\{ - \int_{\varepsilon}^{\infty} \dd\omega  \frac{\dd{I}_{_\R}}{\dd\omega} \right\} = 
\frac{\partial}{\partial \varepsilon} \, 
\exp \left\{ - \int_{\varepsilon}^{\infty} \dd\omega  \frac{\dd{I}_{_\R}}{\dd\omega} \right\} 
\equiv \frac{1}{E} \, \hat{\cal P}_{_\R} \left(\frac{\varepsilon}{E} \right) \, .
\ee 
Since the spectrum \eq{dIR} depends on the induced radiation and incoming parton energies 
($\omega$ and $E$, respectively, in the target rest frame) only through the ratio $\omega/E$, 
the function $\hat{\cal P}_{_\R}$ in \eq{qw-R} is a scaling function of the fractional 
energy loss $x \equiv \varepsilon/E$. Using \eq{dIR} this function can be expressed as
\be
\label{quenching-Spence}
\Phat_{_\R} \left(x, \ell_{_{\perp \rm A}}, M_{\xi} \right) 
=  
\frac{\partial}{\partial x} \, \exp \left\{C_\R \, \frac{\alpha_s}{2 \pi} 
\left[ {\rm Li}_2\left(\frac{-\ell_{_{\perp \rm A}}^2}{x^2 M_{\xi}^2} \right) 
     - {\rm Li}_2\left(\frac{-\ell_{_{\perp \rm p}}^2}{x^2 M_{\xi}^2} \right) \right] \right\}  \, ,
\ee
where ${\rm Li}_2(u) = - \int_0^u \frac{\dd v}{v} \ln(1-v)$ is the Spence function, 
and the dependence of $\Phat_{_\R}$ on $\ell_{_{\perp \rm A}}$ and $M_{\xi}$ is made explicit. 
Note that for a singlet $Q \bar{Q}$ pair, $\Phat_{_{\R}} \to \delta(x)$, 
corresponding to the absence of FCEL in this case.

The transverse momentum broadening $\ell_{_{\perp \rm A}}$ is related to the average 
path length $L_{_{\rm A}}$ in the target nucleus as
\be
\label{gluon-broad}
\ell_{_{\perp \rm A}}^2 = \qhat  L_{_{\rm A}} \, , 
\ee
where $\qhat$ is the transport coefficient in cold nuclear matter parametrised by~\cite{Arleo:2020hat}
\be
\label{qhat-x}
\hat{q} \equiv \hat{q}_{_0} \left( \frac{10^{-2}}{\min(x_{_0}, \xtwo)} \right)^{0.3}  ; 
\ \  x_{_0} = \frac{1}{2 m_\mathrm{p} L_{_{\rm A}}} \ ; 
\ \  \xtwo \sim \frac{2 m_\perp}{\sqrt{s}} \,e^{-y} \, .
\ee
Here 
$m_\mathrm{p}$ is the proton mass, and 
the normalisation parameter $\hat{q}_{_0}$ will be taken as $\hat{q}_{_0} = 0.07\pm0.02$ GeV$^2$/fm, 
as previously estimated from various phenomenological studies~\cite{Arleo:2020hat}. 
The average path length will be set to $L_{_\textnormal{Pb}}=10.11$~fm for a lead nucleus, 
and $L_{_\textnormal{p}}=1.5$~fm for a proton target~\cite{Arleo:2012rs}. 

As a result of the scaling of $\hat{\cal P}$ in $x$, the induced energy loss appearing when going 
from \pp to \pA collisions is naturally accounted for by an energy rescaling, 
or equivalently by a rapidity shift
\be
\label{delta-shift}
\delta = \ln{(1+x)} \, . 
\ee
Within the PDA, FCEL leaves the dijet internal structure unchanged and thus does not alter the 
$Q \bar{Q}$ pair's colour state, rapidity difference $\Delta y$ or invariant mass $M_{\xi}$ 
(in particular, $\xi$ and $K_\perp$ are conserved, cf.~Eq.~\eq{dijet-mass}). 
Hence, the same rapidity shift $\delta$ applies to the pointlike $Q \bar{Q}$ pair, 
its constituents, and in turn to the tagged heavy meson $H$.  

As a consequence, FCEL can be accounted for by relating the heavy meson differential production 
cross sections in \pp and \pA collisions as follows (with $H=D$ or $B$)~\cite{Arleo:2020hat}, 
\be
\label{sig-pA-y} 
\frac{1}{A} \, \frac{\dd \sigma_{\rm pA}^{H}(y,\pt,\sqrt{s})}{\dd y \, \dd \pt}  =  
\sum_{\R} \, \rho_{_{\R}}(\bar{\xi}) \int_0^{x_{\rm max}} \! \! \frac{\dd{x}}{1+x} 
\, \, \Phat_{_\R}(x, \ell_{_{\perp \rm A}}, M_{\bar{\xi}}) 
\, \, \frac{\dd\sigma_{\pp}^{H}(y+\delta, \pt, \sqrt{s})}{\dd y \, \dd \pt} \, .
\ee
Since $\Phat_{_\R}$ depends on the colour state $\R$, 
the rapidity shift $\delta = \ln{(1+x)}$ in \eq{sig-pA-y} is made separately for each $\R$. 
This requires introducing the probability $\rho_{_\R}$ for the dijet to be in colour state $\R$, 
which is determined from the $gg \to Q \bar{Q}$ scattering amplitude and is a function of $\xi$ only, 
$\rho_{_{\R}} = \rho_{_{\R}}(\xi)$, see Appendix \ref{app:colour-proba}. 
The parameter $\bar{\xi}$ in Eq.~\eq{sig-pA-y} can be viewed as the typical $\xi$ in the 
\pp cross section $\dd\sigma_{\pp}^{H}/\dd y \, \dd \pt$. 
Following Ref.~\cite{Arleo:2020hat}, 
the uncertainty associated to the value of $\bar{\xi}$ will be estimated by varying $\bar{\xi}$ 
in the interval $[0.25,0.75]$. Note that $\delta$, $x$, $\xi$ 
(and thus also $\bar{\xi}\,$) are invariant under longitudinal boosts, 
and \eq{sig-pA-y} can thus be equally used in the target rest frame or center-of-mass frame, 
with $y$ the heavy meson rapidity in the chosen frame. 
Finally, in \eq{sig-pA-y} we set $K_{_\perp} = \pt/z$ in the expression of~$M_{\bar{\xi}}$ 
(see Eq.~\eq{dijet-mass}) 
to account for the rescaling of momenta in $Q \to H$ fragmentation 
(the fragmentation variable $z$ will be treated as a parameter, see Sec.~\ref{sec:predictions}), 
and $x_{\rm max} = 1$ for consistency with the soft radiation approximation.\footnote{%
  In general, one should also impose energy conservation, 
  $x \leq (E_{\mathrm{p}} -E)/E$ (with $E_\mathrm{p}$ the projectile proton energy). 
  However, this constraint starts to play a role only at very large $y$, 
  and affects negligibly the integral \eq{sig-pA-y} 
  in the rapidity range considered in the present study.
}

Starting from a given heavy meson production cross section in pp collisions, 
the expression \eq{sig-pA-y} singles out the effect of FCEL on the corresponding pA cross section, 
within the pointlike dijet approximation. 
The pp cross section will be parametrised by Eq.~\eq{eq:fit}, where the overall 
normalisation factor ${\cal N}(\pt)$ is irrelevant for evaluating the ratio \eq{RpA-1}. 

\section{FCEL baseline predictions}
\label{sec:predictions}

Here we provide our main results, based on \eq{sig-pA-y}, 
for heavy meson nuclear suppression expected from FCEL, 
and compare them with available LHC pPb collision data. 
Let us stress that the above implementation of FCEL can be justified, 
since the typical $x$ contributing to \eq{sig-pA-y} turns out to be consistent with the 
PDA \eq{log-accuracy}. 
Defining the typical $x$ as the median $x$ in the integral \eq{sig-pA-y}, 
we have indeed checked that $\ln{(\ell_{_{\perp \rm A}}^2/x^2 m_{_\perp}^2)} \sim 3$--$4.5$ 
(depending on the values of $y$ and $\pt$) for all observables considered in what follows.

\subsection{Observable and parameters}
\label{sec:observables}

The heavy meson \pA cross section \eq{sig-pA-y} will be evaluated using the \pp cross section 
(parametrised like in previous studies to fit the available \pp data, see Appendix~\ref{app-ppfits}) 
as input, and the theoretical prediction for the quenching weight $\hat{\cal P}$ arising from FCEL. 
In other words, we will predict the heavy meson nuclear modification factor 
(in minimum bias \pA collisions compared with \pp collisions), 
\be
\label{RpA-1}
R_{\pA}^{H}\big(\,y,\,\pt,\sqrt{s}\,\big)
  = 
  \frac{1}{A} \, {\frac{\dd\sigma_{\pA}^{H}}{\dd y \, \dd \pt} 
  \biggr/ 
  \frac{\dd\sigma_{\pp}^{H}}{\dd y \, \dd \pt}}  \, , 
\ee
expected solely from the FCEL effect. 

The values of the 
fragmentation variable $z$ (based on the fragmentation functions of Ref.~\cite{Peterson:1982ak}),
exponent parameter $n$ (see Eq.~\eq{eq:fit}), 
and mass $m$ are as indicated in Table~\ref{table:1} for both charm and bottom.
As already mentioned, 
the parameters $\hat{q}_{_0}$ and ${\bar \xi}$ will be taken as 
$\hat{q}_{_0} = 0.07\pm0.02$ GeV$^2$/fm and $\bar{\xi} = 0.50 \pm 0.25$ (for both $D$ and $B$ production). 

\begin{table}[t]
\begin{center}
\begin{tabular}[]{c|ccc}
  \quad meson \quad & $z$ & $n$ & $m$  \\
  \hline
  \quad $D$         \quad & \ $0.8\pm0.2$       & \ $4\pm1$       & \ $1.3\pm0.2$ GeV     \\
  \quad $B$         \quad & \ $0.9\pm0.1$       & \ $2.0\pm0.5$  & \ $4.6\pm0.5$ GeV  \\
\end{tabular}
\caption{
  Parameters $z$, $n$, $m$ and their variations for $D$ and $B$ production. 
  (Note that $m$ is the on-shell mass of the heavy quark that eventually fragments, 
  not the meson's mass. For the $b$-quark mass, we adopt the value from 
  the so-called `1S scheme' \cite{PDG}.)
}
\label{table:1}
\end{center}
\end{table}

Theoretical uncertainties will be estimated as in our previous FCEL studies 
of light hadron~\cite{Arleo:2020hat,Arleo:2020eia} and quarkonium~\cite{Arleo:2014oha} production, 
assuming the parameters to be uncorrelated and determining the uncertainty band of our 
predictions using the Hessian method~\cite{Pumplin:2001ct}. 
The quark was massless in the light hadron studies, here its mass joins the full set of parameters 
$\{\hat{q}_{_0},\bar{\xi},z,n,m\}\,$ in characterising a given prediction.

\subsection{Results for $D$ and $B$ mesons}
\label{sec:results}

The \pA cross section \eq{sig-pA-y} is differential w.r.t.~$y$ and $\pt$, 
both kinematic variables being necessary to specify the FCEL quenching weight. 
Measurements of the ensuing `doubly differential' heavy meson suppression \eq{RpA-1} were 
taken by the LHCb experiment at forward and backward rapidities in the (combined) 
range $2 < |y| < 4.5\,$. The ALICE experiment has measured the $\pt\!$-distribution at midrapidity.

\begin{figure}[t]
\centering
\vskip -8mm 
\includegraphics[scale=\figscale]{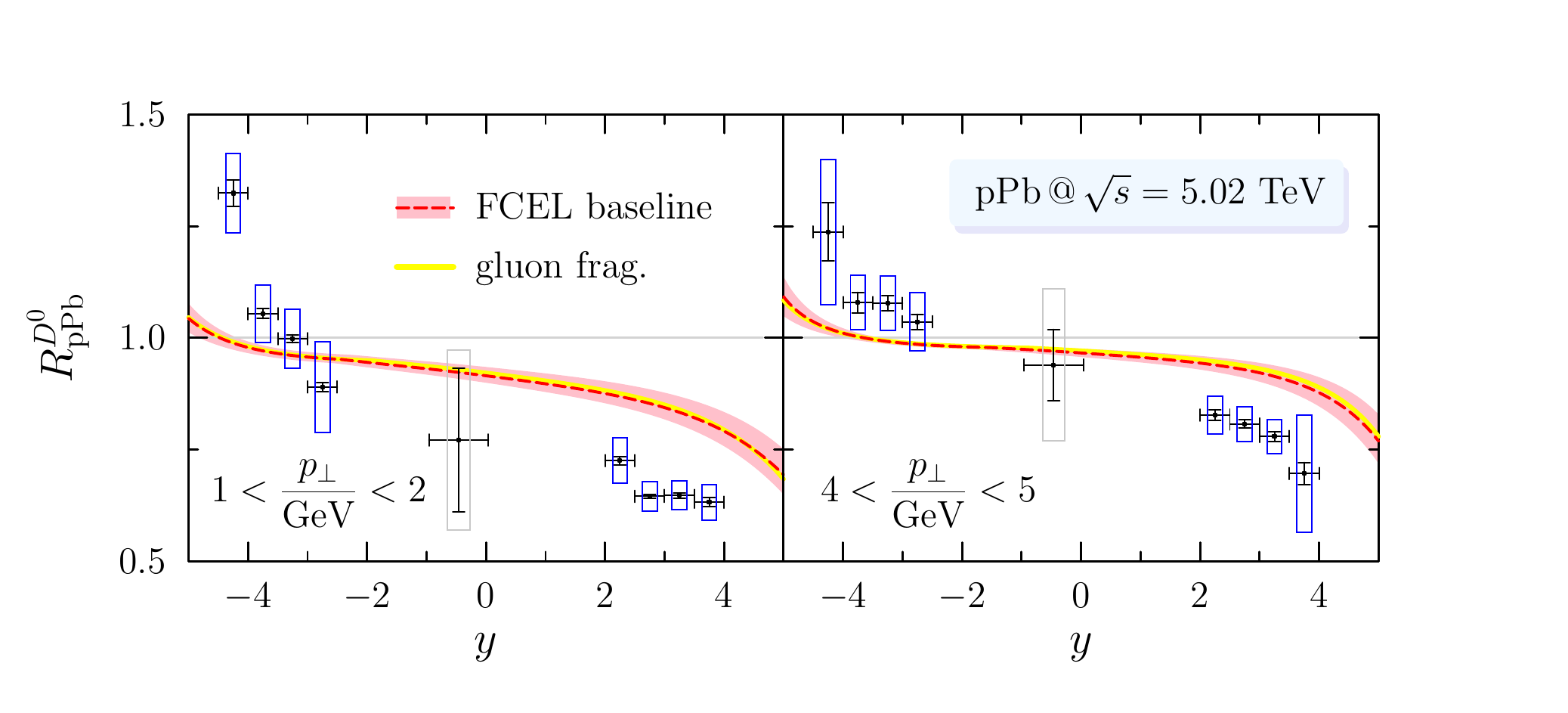}  
\vskip -5mm 
\caption{
  The rapidity dependence of $R_{\rm pPb}$ for $D^0$ production at $\sqrt{s} = 5.02$~TeV 
  is shown for two $\pt\!$-bins. The baseline calculation is for the $gg \to Q {\bar Q}$ 
  channel in the center of each bin, compared with ALICE and LHCb 
  data~\cite{Abelev:2014hha,Aaij:2017gcy}. 
  (The curves labelled `gluon frag.' correspond to the channel $gg \to gG \to gQ {\bar Q}$ 
  discussed in Sec.~\ref{sec:nlo}.)
}
\vskip 5mm
\label{fig-RPA-D}
\end{figure}

For $D^0$ production at $\sqrt{s}=5.02$ TeV, we display in Fig.~\ref{fig-RPA-D} data from 
Refs.~\cite{Abelev:2014hha,Aaij:2017gcy} alongside the FCEL results as a function of $y$, 
for two $\pt\!$-bins.\footnote{\label{panorama-plots}
  A complete comparison, with plots for all $\pt\!$-bins (with $\pt < 10$ GeV), 
  is included as an ancillary file in the arXiv record.
}
As in previous studies of quarkonium~\cite{Arleo:2012hn,Arleo:2012rs,Arleo:2013zua} 
and light hadron production~\cite{Arleo:2020eia,Arleo:2020hat}, 
the increase of the suppression with increasing $y$ is a direct consequence of the 
scaling of FCEL with the incoming parton energy.
We observe a good agreement between this overall trend and the data, 
which also holds for other $\pt\!$-bins. 
The chosen $\pt\!$-bins also demonstrate another important feature of FCEL, namely, 
it becomes weaker at larger $\pt$ 
(as illustrated by the parametric dependence \eq{aveFCEL} of the average FCEL). 

Our main message is already clear from Fig.~\ref{fig-RPA-D}: the FCEL effect by itself explains 
about 
half of the nuclear suppression of $D^0$ mesons at forward rapidities (for those $\pt\!$-bins). 
Note that as a purely perturbative effect, 
FCEL is not predicted to distinguish between neutral and charged mesons.

\begin{figure}[t]
\centering
\hskip -5mm \includegraphics[scale=0.6]{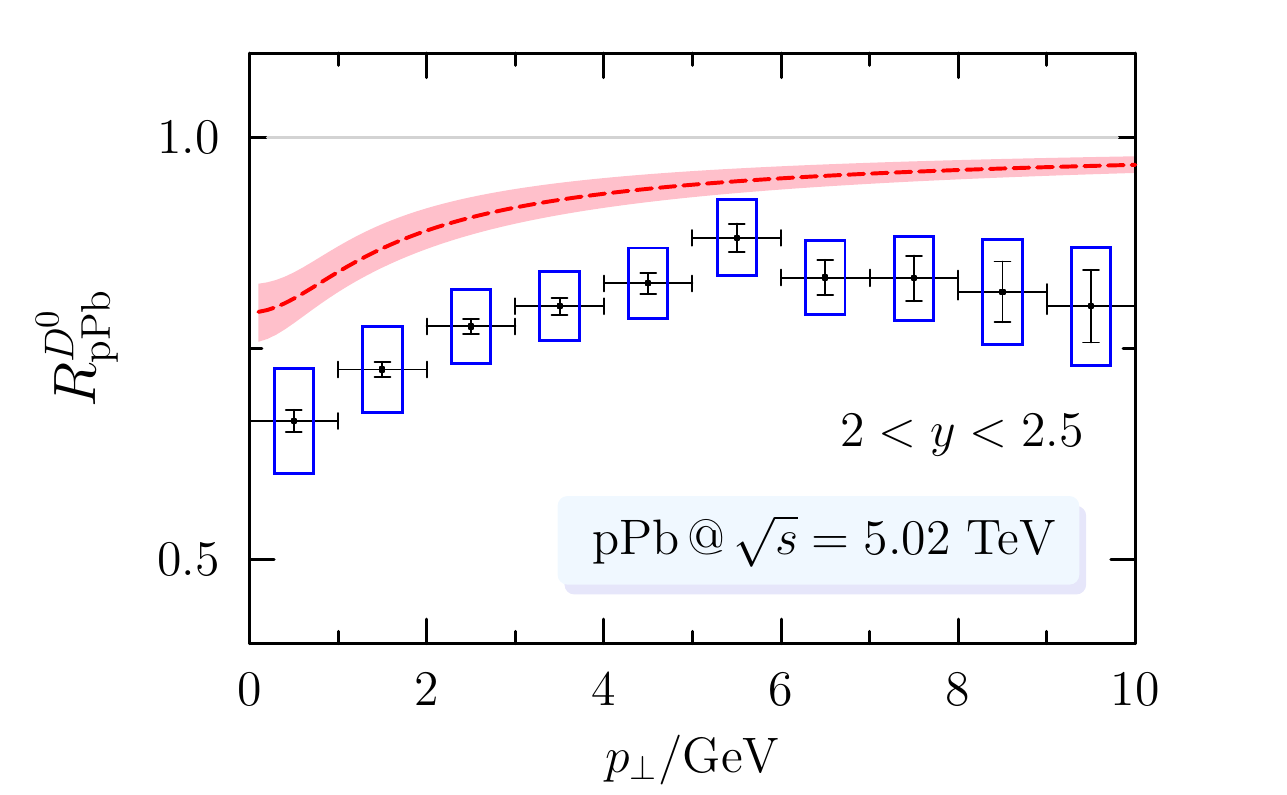} 
\hskip -5mm \includegraphics[scale=0.6]{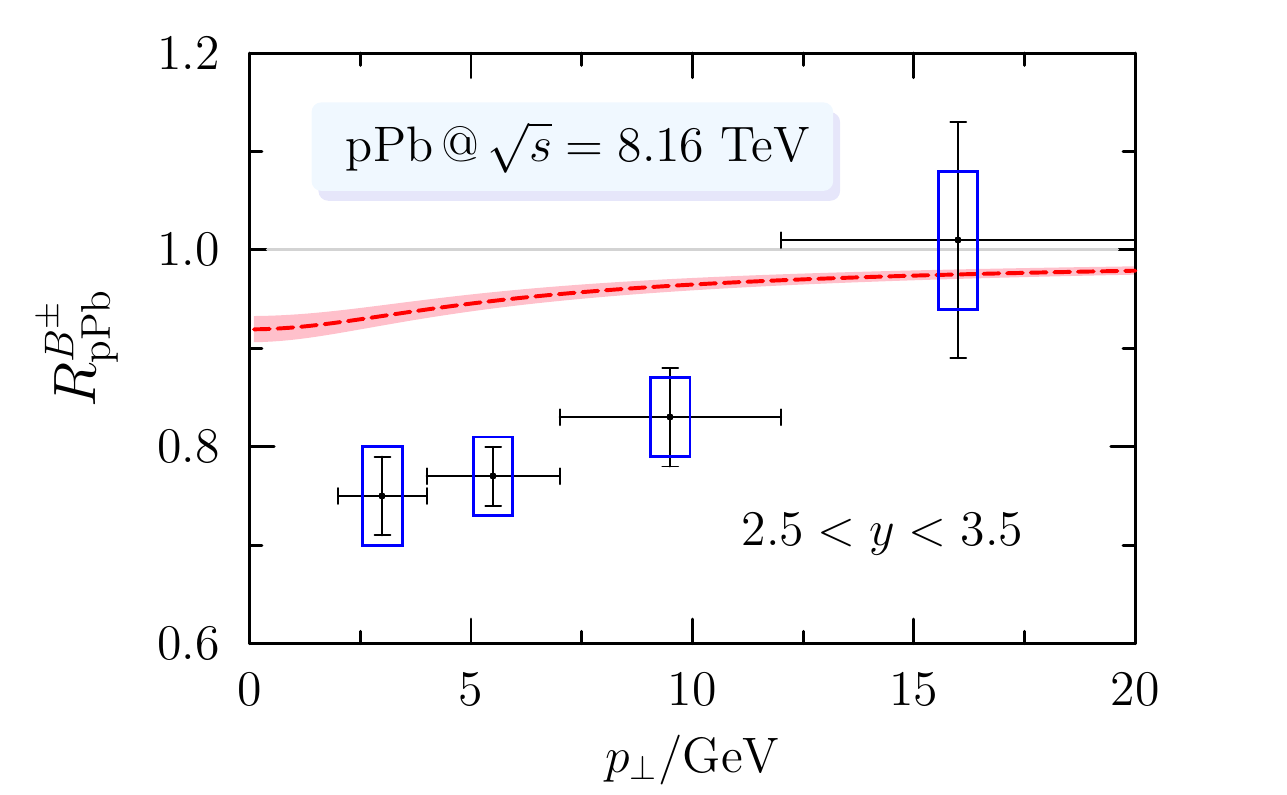}  
\caption{
  Comparison between LHCb data~\cite{Aaij:2017gcy,Aaij:2019lkm} and FCEL baseline 
  predictions of the $\pt\!$-dependence of $R_{\rm pPb}$, for $D^0$ production at 
  $\sqrt{s} = 5.02$~TeV in the rapidity interval $2<y<2.5\,$ (left), 
  and for $B^\pm$ production at $\sqrt{s} = 8.16$~TeV in the interval $2.5<y<3.5$ (right).
}
\label{fig-RPA-2}
\end{figure}

In Fig.~\ref{fig-RPA-2} (left) we show the same data as a function of $\pt$, 
for $2 < y < 2.5$, together with our FCEL baseline prediction evaluated at $y=2.25\,$.\footnote{%
  \label{panorama-plots-y}Plots for all $y$-bins can be found in the ancillary files of the arXiv record.
}
The fact that the FCEL effect (\ie, the deviation of $R_{\pA}^{H}$ w.r.t.~unity) 
decreases with increasing $\pt$ is more visible on this plot. 
The importance of the FCEL effect is also evident on Fig.~\ref{fig-RPA-2} (left): 
for the chosen $y$-interval, it explains 
half of the $D^0$ nuclear suppression up to $\pt \sim 5$~GeV. 
Since LHCb has also measured $B$-meson suppression in pPb collisions at 
$\sqrt{s}=8.16$~TeV~\cite{Aaij:2019lkm}, 
a similar plot for $B^\pm$ production is provided in Fig.~\ref{fig-RPA-2} (right), 
over a larger rapidity window ($2.5<y<3.5$) and for $\pt \leq 20$~GeV.

\subsection{Influence of NLO processes -- a rough estimate}
\label{sec:nlo}

The results presented in the previous sections assumed the LO picture described in Sec.~\ref{sec:setup}. 
In particular, FCEL has been implemented assuming that the pp cross section is dominated by 
the LO process $gg \to Q\bar{Q}$ 
(which is indeed dominant over other LO processes for $Q\bar{Q}$ production at LHC energies).

On the other hand, we have taken the pp cross section to fit the pp data, 
which are likely to receive an important contribution from higher-order processes. 
Clearly, a full NLO calculation implementing FCEL is beyond the scope of our study. 
In order to estimate the uncertainty associated with the LO picture used in our model, 
we investigate FCEL effects in a generic channel contributing to the pp cross section at NLO 
(in the fixed flavour number scheme adopted here, see Sec.~\ref{sec:setup}), namely, $gg \to gG$ 
(where $G$ denotes a `massive gluon' of mass $2m$) followed by collinear gluon fragmentation 
$G \to Q\bar{Q}$ ($Q$ and $\bar{Q}$ sharing equally the momentum of gluon $G$). 

In the approximation where the induced radiation does not resolve the final parton system 
(denoted as PDA in Sec.~\ref{sec:rescaling}), FCEL depends on the partonic subprocess only through the 
invariant mass and colour probabilities $\rho_{_\R}$ of that system. 
In order to estimate FCEL effects assuming the above $gg \to g G \to g Q\bar{Q}$ process, 
we can thus use \eq{sig-pA-y}, up to the following modifications: 
(i) the mass $M_{\xi}$ of the gluon pair produced in $gg \to gG$ is now given by 
$M_{\xi}^2 = {K_{_\perp}^2}/{\bm(\xi(1-\xi)\bm)} + {4m^2}/{\xi}$ 
(with $\xi$ the energy fraction carried by gluon $G$, and $K_{_\perp} = 2 \pt/z$); 
(ii) the sum over $\R$ runs over three colour states, $\R = {\bf 1}, {\bf 8}, {\bf 27}$, 
with associated probabilities given in \eq{colour-proba-pheno-gg}.

Repeating the FCEL calculation of Sec.~\ref{sec:results} for heavy meson nuclear suppression 
with the above modifications, we expect two competing effects. 
The larger `dijet mass' should entail a milder  suppression, whereas the richer colour structure 
(in particular, the presence of the larger Casimir $C_{{\bf 27}}=8$) 
should on the contrary strengthen the suppression. 
For $D^0$ suppression as a function of $y$, we obtained the yellow curves in Fig.~\ref{fig-RPA-D}. 
The $gg\to gG$ process appears to coincide with the $gg \to Q{\bar Q}$ LO baseline, 
confirming the expected partial compensation between the two competing effects. 

Although this estimate of the possible effect of NLO processes on our predictions is rudimentary, 
it makes us confident that the results obtained within the LO picture are quantitatively meaningful.
We expect the main feature, namely, an increase of both the dijet mass and average Casimir, 
to prevail when going from LO to NLO processes. 

\section{Discussion}
\label{sec:discussion}

As noted in Ref.~\cite{Arleo:2010rb}, 
where the fully coherent medium-induced gluon radiation spectrum has first been derived, 
FCEL is expected to affect \emph{all} hadron species in proton-nucleus collisions. 
In the present study, we apply it to the case of open heavy-flavour hadron production. 
The results show that FCEL is a sizable effect, accounting for about half of the $D$-meson 
nuclear suppression observed at forward rapidity, in a wide $\pt\!$-range. 
After studies on quarkonium and light hadron production, 
this confirms that hadron production in pA collisions cannot be 
described within the collinear factorization approach using only nPDFs, calling for a change of paradigm. 

Nevertheless, it has been recently suggested to use the data on heavy-flavour hadron production 
in pA collisions as a reliable probe of gluon distributions in nuclei, assuming nPDFs to be 
the only nuclear effect at work~\cite{Kusina:2017gkz,Eskola:2019bgf,Kusina:2020dki}. 
The latter claim is 
based on the relatively good agreement of pQCD calculations with heavy-flavour measurements 
in pPb collisions at the LHC, after a proper reweighting of nPDFs using precisely these data sets. 
Although such an agreement between data and theory (after reweighting) is necessary to justify 
the use of collinear factorization in pA collisions, it should not be seen as a sufficient condition, 
let alone a proof of the absence of parton dynamics beyond collinear factorization. 
The results shown here indeed demonstrate that a significant part of the suppression 
observed in data is due to FCEL, 
an effect which breaks explicitly factorization. 
Our claim is supported by the precision of the FCEL calculation, 
with the moderate relative theoretical uncertainty on $R_{\pA}^H$ (typically below $10$\%) 
ensuing from FCEL being fully determined within pQCD. 

FCEL does not affect the rate of hard processes in nuclear collisions in the same way as nPDFs do 
--~for instance as a function of $\sqrt{s}$, $M$, $\pt$, or $y$~-- due to the different scaling 
properties of these two nuclear effects. 
To illustrate this, the effects of FCEL scale approximately 
as the momentum fraction $x_1$ carried by the parton in the hadron projectile\footnote{%
  This scaling is slightly violated because of the energy evolution of the transport coefficient, 
  see Ref.~\cite{Arleo:2012rs}.
} 
while nPDF effects are expected to depend on $x_2$, 
the momentum fraction carried by the nuclear target parton. 
Moreover, the parametric dependence on the hard scale should be different for both effects.
In particular, the {\it average} FCEL associated to the spectrum \eq{dIR} is suppressed 
by one power of the transverse mass~\cite{Arleo:2010rb,Peigne:2014uha}, 
\be
\label{aveFCEL}
\Delta E_{_{\rm FCEL}} 
\propto \alpha_s \, \frac{\ell_{_{\perp\rm A}} - \ell_{_{\perp\rm p}}}{m_\perp} \, E \, .
\ee

Consequently, using the measurements of open heavy-flavour meson production 
(as well as light hadron and quarkonium production) 
in a global fit analysis that ignores the reality of FCEL would lead to an 
incorrect determination of nuclear parton distributions, independently of the apparent 
agreement reached between data and theory. For instance, 
it is shown in Ref.~\cite{Eskola:2019bgf} that including LHCb $D$-meson data has a spectacular 
impact on the determination of 
(reweighted) EPPS16 nPDF sets~\cite{Eskola:2016oht}. 
Not only does it lead to a stronger gluon shadowing of the central set at small values of $x$,\footnote{
  This is reminiscent of the use of forward light-hadron measurements by BRAHMS in dAu collisions at 
  RHIC~\cite{Arsene:2004ux} in the EPS08 analysis~\cite{Eskola:2008ca}, 
  which led to strong nuclear shadowing. 
  These data were subsequently left out in the global fit analysis of EPS09~\cite{Eskola:2009uj}, 
  resulting in a milder gluon shadowing.
} 
but the uncertainty of the reweighted EPPS16 shrinks dramatically, especially at small resolution scales. 
This does not come as a surprise because the central set of (default) EPPS16 tends to 
overshoot LHCb data at forward rapidity, 
and because the precision of LHCb measurements exceeds by far that of EPPS16. 
However, we believe that these reweighted nPDF sets should not be trusted because of the wrong 
physical hypothesis, namely assuming that $D$-meson suppression is only driven by nPDFs. 
The fact that FCEL accounts for half of the suppression at forward rapidity is likely to 
lead to the opposite conclusion of {\it lesser} gluon shadowing at 
small $x$ than in the default nPDF sets. 
The inclusion of forward prompt and non-prompt $J/\psi$ measurements 
in the determination of new nPDF sets, as advocated in~\cite{Kusina:2020dki}, 
would lead to similar biases.

Ideally, the nPDF global fit analyses should include data which are insensitive to FCEL, 
such as DIS measurements or weak boson production in pA collisions, 
or barely affected by FCEL, \eg, jet production at very large $\pt^2 \gg \qhat L$. 
However, the constraints will be much looser especially in the gluon sector and at small $x$. 
Another way would be to use the reweighting method~\cite{Giele:1998gw,Ball:2011gg}, 
which up to now has been used to iteratively include newly available data into
existing nPDF sets (without having to redo the full analysis). We 
propose 
to reweight nPDF sets with a fitting procedure that takes into account new (theoretical) 
information: namely, the inclusion of FCEL for hadron production in pA collisions. 
Because the FCEL uncertainties are relatively 
narrow, 
treating both nuclear effects together should help assess the current tensions and improve 
the overall precision of nPDFs. This programme is left for future work.

\acknowledgments
This work is funded by the ``Agence Nationale de la Recherche'' 
under grant ANR-COLDLOSS (ANR-18-CE31-0024-02). 
G.~J. is funded by the U.S. Department of Energy (DOE) under grant No.~DE-FG02-00ER41132. 

\appendix

\section{Colour state probabilities $\bm{\rho_{_{\rm R}}(\xi)}$}
\label{app:colour-proba}

In this Appendix we present a simple derivation of the probabilities $\rho_{_\R}(\xi)$ 
for the $Q \bar{Q}$ pair produced in $g g \to Q \bar{Q}$ to be in colour state $\R$. 
These probabilities turn out to depend only on the light-cone momentum fraction $\xi \equiv K^+/p^+$ 
(with $K^+$ and $p^+$ the light-cone momenta of the heavy quark and incoming projectile gluon, 
respectively), which can be viewed as an energy fraction in the target rest frame. 
In particular, the colour probabilities are independent of the heavy quark mass $m$. 

Let the target gluon carry momentum $q$ and Lorentz index $\mu$. 
The $g g \to Q \bar{Q}$ scattering amplitude ${\cal M}_{\rm hard}^{\mu}$ can be 
conveniently calculated from the $g \to Q \bar{Q}$ forward scattering amplitude off an 
external gluon field. 
Indeed, in the high-energy limit ($p^+ \to \infty$), 
the latter selects the $\mu=+$ component of ${\cal M}_{\rm hard}^{\mu}$, which can be easily 
derived using light-cone perturbation theory~\cite{Lepage:1980fj} in light-cone $A^+ =0$ gauge. 
The other (dominantly transverse) components of ${\cal M}_{\rm hard}^{\mu}$ are simply obtained 
using gauge invariance, $q_{\mu} {\cal M}_{\rm hard}^{\mu} = 0$, 
and ${\cal M}_{\rm hard}^{\mu}$ thus directly follows from the 
knowledge of ${\cal M}_{\rm hard}^{+}$ only. We find
\be
\label{g2QQbar-amp}
{\cal M}_{\rm hard}^{+} = 
2 g_s  \left\{ \psi(\xi,\Kvec) \, \GBuqqbar(14,-47) + \psi(\xi,\Kvec - \qvec) \, 
\GBtqqbar(14,-45) - \psi(\xi,\Kvec - \xi \qvec) \,  \GBsqqbar(14,-45) \right\} \, ,
\ee
where the graphs stand for the colour factors associated to each Feynman diagram,
$\qvec$ and $\Kvec$ are the transverse momenta of the target gluon and heavy quark, respectively, 
$\psi(\xi,\Kvec)$ is the $g \to Q \bar{Q}$ light-cone wavefunction,\footnote{
  It is given by $\psi(\xi,\Kvec) = g_s \, p^+ \sqrt{\xi(1-\xi)} \, \hat{V} \ln{(\Kvec^2+m^2)}$, 
  where 
  $\hat{V} = \delta_{\sigma}^{-\sigma'} (\xi-\delta_{\sigma}^{-\lambda}) 
  \, {\boldsymbol \varepsilon}_{\lambda} \cdot {\boldsymbol \nabla}_{\Kvec} 
  - \frac{1}{\sqrt{2}} \delta_{\sigma}^{\sigma'} \delta_{\sigma}^{\lambda} 
  \frac{\partial}{\partial m}$, 
  with $\sigma$ and $\sigma'$ denoting the quark and antiquark helicities, 
  respectively, and ${\boldsymbol \varepsilon}_{\lambda} = -\frac{1}{\sqrt{2}} (\lambda, i)$ 
  the transverse polarization ($\lambda=\pm$) of the energetic gluon of light-cone momentum $p^+$. 
  The precise form of $\psi(\xi,\Kvec)$ is however irrelevant to the present discussion.
}
and $g_s = \sqrt{4\pi \alpha_s}$. 

In the limit $q_\perp \ll K_\perp$ considered in the present study 
(recall that $K_\perp \gg \ell_{_{\perp \rm A}} = \sqrt{\qhat  L_{_{\rm A}}}$, 
cf. Sec.~\ref{sec:setup}, and typically $q_\perp \sim \Lambda_{\rm QCD}$) 
the Taylor expansion of \eq{g2QQbar-amp} yields
\be
\label{g2QQbar-amp-Taylor}
{\cal M}_{\rm hard}^{+} = 
- 2 g_s  \left( \qvec \cdot {\boldsymbol \nabla}_{_\Kvec}  \psi(\xi,\Kvec) \right) 
\left\{  \GBtqqbar(14,-45) - \xi \,   \GBsqqbar(14,-45) \right\} \, ,
\ee
where we used colour conservation: 
\be
\GBuqqbar(14,-47) +  \GBtqqbar(14,-45) - \GBsqqbar(14,-45) = 0 \, .
\ee

The dependence of ${\cal M}_{\rm hard}^{+}$ on the heavy quark mass $m$ 
(contained in the first factor of Eq.~\eq{g2QQbar-amp-Taylor}) and its colour structure 
(second factor of Eq.~\eq{g2QQbar-amp-Taylor}) fully factorise. 
As a consequence, the mass dependence cancels out in the colour probabilities 
$\rho_{_\R}$ for the $Q \bar{Q}$ pair to be in colour state $\R$ (with $\R = {\bf 1}, \, {\bf 8}$) 
defined by
\be
\label{proba-def}
\rho_{_{\R}} = \frac{|{\cal M}_{\rm hard} \cdot \mathds{P}_{_\R}|^2}{|{\cal M}_{\rm hard}|^2} \, ,
\ee
where $\mathds{P}_{_\R}$ is the hermitian projector on the colour state $\R$. 
Those probabilities thus coincide with those obtained in Ref.~\cite{Arleo:2020hat} 
for $g g \to q \bar{q}$ with massless quarks, namely, 
\be
\label{colour-proba-qqbar}
\rho_{_{\bf 1}}(\xi) = \frac{1}{9 \bm(\xi^2 + (1-\xi)^2\bm) -1}  \ ; 
\ \ \rho_{_{\bf 8}}(\xi) = 1 - \rho_{_{\bf 1}}(\xi) \, .
\ee

The above discussion applies similarly to the $g g \to gG$ channel (see Sec.~\ref{sec:nlo}) 
and its associated colour states 
(${\bf 8 \otimes 8 = 1 \oplus 8_a \oplus 8_s \oplus 10  \oplus \overline{10} \oplus 27}$), 
resulting in the same colour probabilities as for $g g \to g g$ \cite{Arleo:2020hat} 
\be
\label{colour-proba-pheno-gg}
\rho_{_{\bf 27}}^{gg}(\xi) = 
\frac{3/4}{1+ \xi^2 + (1-\xi)^2}\ ; 
\ \ \rho_{_{\bf 1}}^{gg}(\xi) = \frac{1}{3} \rho_{_{\bf 27}}^{gg}(\xi) \ ; 
\ \ \rho_{_{\bf  8}}^{gg}(\xi) = 1 -  \frac{4}{3} \rho_{_{\bf 27}}^{gg}(\xi) \ ; 
\ \ \rho_{_{\bf 10}}^{gg} = 0 \, ,
\ee
where the colour representations with the same dimension and Casimir have been combined.

\begin{figure}[t]
\centering
\vskip -8mm 
\includegraphics[scale=\figscale]{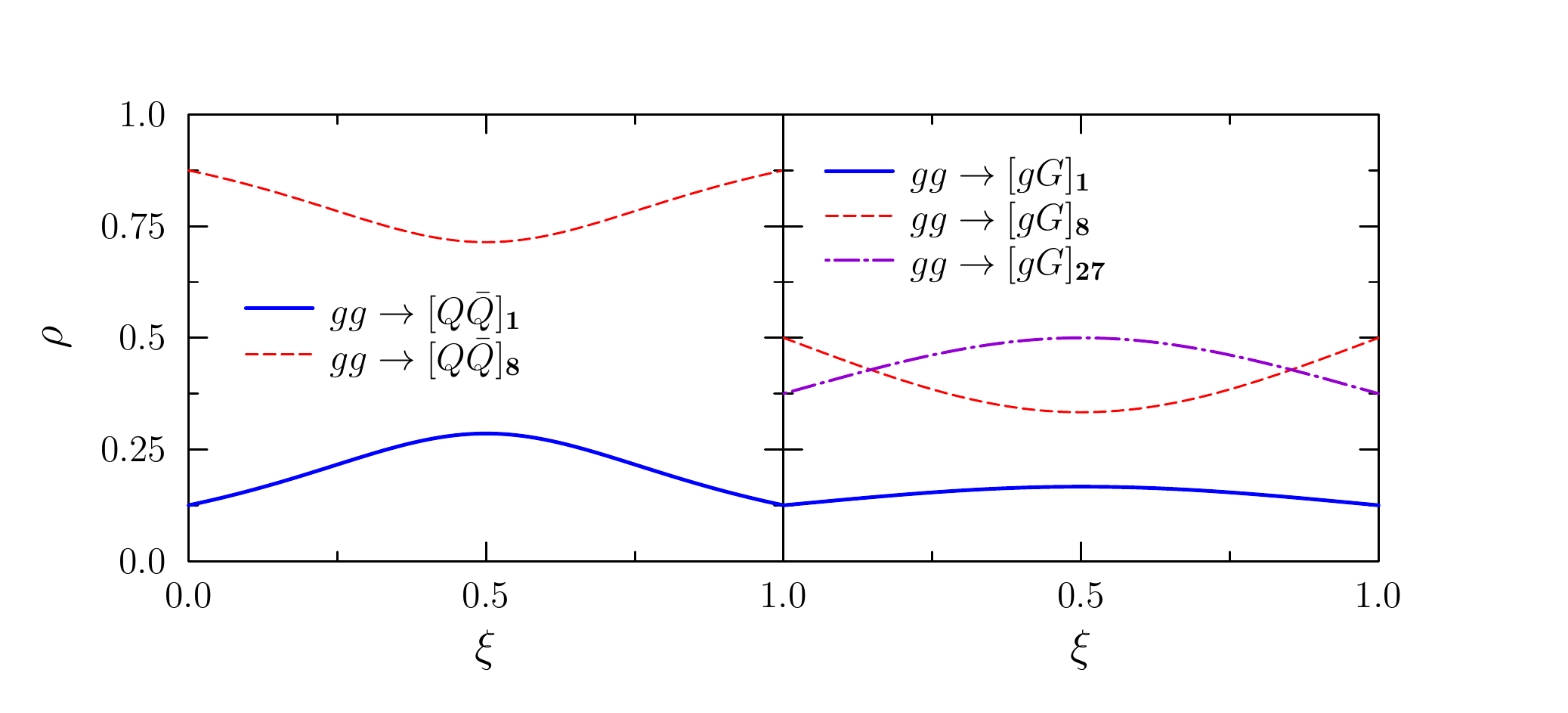}
\caption{
  Probabilities $\rho_{_{\R}}(\xi)$ for the produced dijet to be in colour state $\R$, 
  as a function of the internal energy fraction $\xi$, 
  in the $gg \to Q \bar{Q}$ channel for $\R = {\bf 1}, \, {\bf 8}$ (left) and 
  in the $gg \to gG$ channel for 
  $\R={\bf 1}, \, {\bf 8}\equiv {\bf 8_a \oplus  8_s}, \, {\bf 27}$ (right). 
} 
\label{fig-probas}
\end{figure}

The probabilities from Eqs.~\eq{colour-proba-qqbar} and \eq{colour-proba-pheno-gg} 
are shown in Fig.~\ref{fig-probas}.

\section{Parametrisation of heavy meson cross section in \pp collisions}
\label{app-ppfits}

To make predictions for $R_{\pA}$ based on the model of FCEL in Eq.~\eq{sig-pA-y}, 
a main input is the doubly differential heavy meson production cross section in pp collisions. 
The latter is calculable within pQCD but subject to proton PDFs and fragmentation functions 
\cite{Martin:1998sq,Peterson:1982ak}. 
For our purpose of predicting the 
{\it ratio} \eq{RpA-1} 
rather than absolute pp and pA cross sections, 
we instead use a parametrisation of the pp cross section allowing to fit best the pp data. 

In order to trust the pp cross section for the kinematic regimes 
where our predictions for $R_{\pA}$ are made, 
we adopt the parametrisation
\bea
\label{eq:fit}
\frac{ \dd \sigma_{\rm pp}^H }{\dd y \,\dd \pt } 
\ = \ 
{\cal N}(\pt) \, \Big[ \big(1-\chi\big)\big(1-\sqrt{\chi}\,\big) \Big]^n 
\, , \quad \chi \equiv 4 \left( \frac{\pt^2 +\mu_H^2}{s} \right)^{\! _\frac{1}{2}} \cosh y 
\, . \hskip 5mm && 
\eea
If we let the overall normalisation in 
Eq.~\eq{eq:fit} be treated as a free parameter for each $\pt\!$-bin, 
this parametrisation 
is capable of describing LHCb pp data at 
$\sqrt{s}=\{5, 7, 13\}$~TeV~\cite{Aaij:2016jht,Aaij:2013mga,Aaij:2015bpa,Aaij:2013noa}, 
for both charm and bottom production, with parameters $\mu_D = 1.8$~GeV and 
$n = 4 \pm 1\,$, and $\mu_B = 5.3$~GeV and $n = 2.0 \pm 0.5\,$, respectively.\footnote{
  The fact that totally different values of the `exponent parameter' $n$ were used for 
  light hadron production (namely, $n=15\pm5$)~\cite{Arleo:2020hat} should not lead to confusion, 
  the parametric form \eq{eq:fit} chosen here to fit the heavy meson pp cross section 
  being different from that used in  Ref.~\cite{Arleo:2020hat} to fit the 
  light hadron cross section. 
  The precise form of the parametrization used to fit the pp cross section is irrelevant in our approach.
} 
For simplicity we choose to fix the value of $\mu_H$ close to the meson mass, 
because the variation in the exponent $n$ is more than sufficient to encompass the data.

\begin{figure}[t]
\centering
\includegraphics[scale=\figscale]{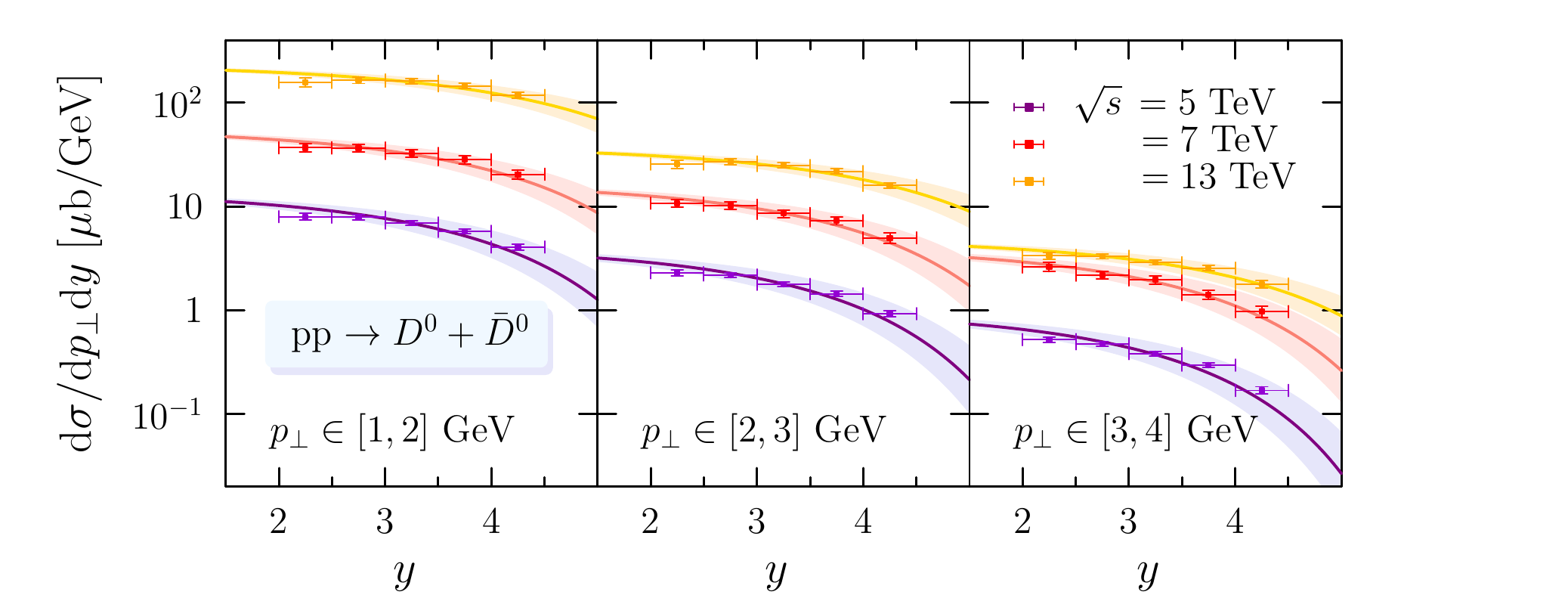}
\caption{
  Differential pp cross section for $D^0$ production as a function of rapidity, 
  at the collision energies $\sqrt{s} = \{5,7,13\}$~TeV. 
  The panels depict \eq{eq:fit} evaluated at the central values of three experimental 
  $\pt\!$-bins, namely, $\pt = 1.5$~GeV (left), $2.5$~GeV (central) and $3.5$~GeV (right). 
  (For the $13$~TeV data, two bins were combined.) 
  The bands correspond to the variation of the parameter $n = 4 \pm 1$ in \eq{eq:fit}. 
  The curves (and data points) at different $\sqrt{s}$ were multiplied by $4^{-k}$ 
  (with $k=0,1,2$ for decreasing $\sqrt{s}$).
}
\label{fig-xspp-pT}
\end{figure}

In Fig.~\ref{fig-xspp-pT}, neutral $D$-meson production from available pp data at forward rapidities 
is compared with the parametrisation for $1$ GeV $<\pt<4$ GeV. Evidently, 
\eq{eq:fit} should be applicable at the intermediate $\sqrt{s} = 8.16$ TeV -- 
despite no corresponding data. The available 
\pp data for charged heavy mesons ($D^\pm$, $B^\pm$) are similarly well 
encompassed by our choices for the exponent $n$, 
for all values of $\pt$ used in this study. 

\bibliography{mybib}
\bibliographystyle{JHEP}

\end{document}

%% file: macro.tex
\def\bc{\begin{center}}
\def\ec{\end{center}}
\def\bi{\begin{itemize}}
\def\ei{\end{itemize}}
\def\be{\begin{equation}}
\def\ee{\end{equation}}
\def\bea{\begin{eqnarray}}
\def\eea{\end{eqnarray}}
 
\newcommand{\ie}{{i.e.}}  
\newcommand{\eg}{{e.g.}}

\def\xtwo{x_{_2}}

\newcommand{\morder}[1]{{\cal O}\left(#1 \right)}
\newcommand{\eq}[1]{(\ref{#1})}

\newcommand{\tf}{t_{\mathrm{f}}}
\newcommand{\R}{\textnormal{R}}
\newcommand{\Kvec}{{\boldsymbol K}}
\newcommand{\qvec}{{\boldsymbol q}}

\newcommand{\pt}{p_{_\perp}}
\newcommand{\pp}{\ensuremath{\text{pp}}\xspace}
\newcommand{\pA}{\ensuremath{\text{pA}}\xspace}

\newcommand{\dd}{{\rm d}}
\newcommand{\lsim}{\lesssim}

\def\bm#1{\mbox{\boldmath$#1$}}

\newcommand{\Phat}{\hat{{\cal P}}}
\def\qhat{\hat{q}}

\def \GenericGtoQQbar  (#1,#2,#3) {
\resizebox{#1 mm}{!}{\raisebox{#2 pt}{
\begin{axopicture}(451,116) (31,-89)
    \SetWidth{1.0}
    \SetColor{Black}
    \Gluon(128.772,-21.27)(45.99,-21.27){2.587}{10}
    \SetWidth{1.0}
    \Line(217.877,-70.134)(223.625,-64.386)\Line(217.877,-64.386)(223.625,-70.134)
    \Line(236.273,-70.134)(242.021,-64.386)\Line(236.273,-64.386)(242.021,-70.134)
    \Gluon(220.751,-67.26)(220.176,-13.222){2.587}{9}
    \Text(465,-53)[l]{\fontsize{18}{2}\selectfont {$H$}}
    \Text(440,-70)[l]{\fontsize{#3}{2}\selectfont {$p_{\perp} = z K_\perp$}}
    \Text(84.506,-6.324)[l]{\fontsize{#3}{2}\selectfont {$E$}}
    \Gluon(257.543,-67.26)(256.968,-8.623){2.587}{9}
    \Line(254.669,-70.134)(260.417,-64.386)\Line(254.669,-64.386)(260.417,-70.134)
    \SetWidth{1.0}
    \Line[arrow,arrowpos=0.35,arrowlength=10,arrowwidth=5,arrowinset=0.2](441.503,15.522)(128.772,-21.27)
    \Text(320,-55)[l]{\fontsize{#3}{2}\selectfont {$\xi, \ \Kvec_1$}}
    \SetWidth{1.0}
    \Line(48.864,-27.594)(62.661,-31.043)
    \Line(48.289,-19.546)(62.661,-14.372)
    \Line(47.714,-16.096)(60.937,-6.898)
    \Line(32.193,-24.145)(47.714,-24.145)
    \Line(32.193,-18.971)(47.714,-18.971)
    \SetWidth{1.0}
    \GOval(47.714,-22.42)(10.923,5.174)(0){0.882}
    \Gluon(239.147,-67.26)(239.147,-34.492){2.587}{5}
    \Text(300,20)[l]{\fontsize{#3}{2}\selectfont {$1-\xi, \ \Kvec_2$}}
    \Oval(239.722,-33.343)(55.188,55.188)(0)
    \SetWidth{1.0}
    \Line[arrow,arrowpos=0.28,arrowlength=10,arrowwidth=5,arrowinset=0.2,flip](413.909,-48.864)(128.772,-21.27)
    \SetWidth{1.0}
    \Line(450.701,-54.038)(463.923,-54.613)
    \Line(421.382,-53.463)(448.401,-55.188)
    \Line(420.232,-49.439)(450.701,-51.739)
    \Line(420.232,-48.864)(436.904,-44.265)
    \Line(418.508,-45.415)(435.754,-37.367)
    \SetWidth{1.0}
    \GOval(416.783,-50.589)(8.048,7.473)(0){0.882}
    \GOval(449.551,-53.463)(4.599,4.599)(0){0.882}
  \end{axopicture}
}}}

\def \GBsqqbar  (#1,#2) {
\resizebox{#1 mm}{!}{\raisebox{#2 pt}{
  \begin{axopicture}(228,113) (222,-131)
    \SetWidth{3.0}
    \SetColor{Black}
    \Line[arrow,arrowpos=0.7,arrowlength=20,arrowwidth=10,arrowinset=0.2](368,-31)(416,-127)
    \Gluon(310,-43)(256,-127){11}{4}
    \Gluon(368,-31)(224,-31){11}{5}
    \Line[arrow,arrowpos=0.5,arrowlength=20,arrowwidth=10,arrowinset=0.2,flip](368,-31)(448,-31)
  \end{axopicture}
}}}

\def \GBuqqbar  (#1,#2) {
\resizebox{#1 mm}{!}{\raisebox{#2 pt}{
 \begin{axopicture}(228,116) (222,-131)
    \SetWidth{3.0}
    \SetColor{Black}
    \Line[arrow,arrowpos=0.75,arrowlength=20,arrowwidth=10,arrowinset=0.2,flip](304,-28)(448,-28)
    \Gluon(384,-28)(256,-124){11}{6}
    \Gluon(304,-28)(224,-28){11}{3}
    \Line[arrow,arrowpos=0.3,arrowlength=20,arrowwidth=10,arrowinset=0.2,flip](416,-124)(304,-28)
  \end{axopicture}
  }}}

\def \GBtqqbar  (#1,#2) {
\resizebox{#1 mm}{!}{\raisebox{#2 pt}{
 \begin{axopicture}(228,116) (222,-131)
    \SetWidth{3.0}
    \SetColor{Black}
    \Line[arrow,arrowpos=0.75,arrowlength=20,arrowwidth=10,arrowinset=0.2,flip](304,-28)(448,-28)
    \Gluon(340,-60)(256,-124){11}{4}
    \Gluon(304,-28)(224,-28){11}{3}
    \Line[arrow,arrowpos=0.3,arrowlength=20,arrowwidth=10,arrowinset=0.2,flip](416,-124)(304,-28)
  \end{axopicture}
  }}}